\documentstyle[prl,preprint,aps,psfig]{revtex}
\begin{document}

\tighten

\title{Optimization of Bell's Inequality Violation For Continuous Variable
Systems }
\author{G.~Gour\thanks{E-mail:~gilgour@phys.ualberta.ca},
F. C.~Khanna\thanks{E-mail:~khanna@phys.ualberta.ca},
A.~Mann\thanks{E-mail:~ady@physics.technion.ac.il},
M.~Revzen\thanks{E-mail:~revzen@physics.technion.ac.il}}
\address{Theoretical Physics Institute,
Department of Physics, University of Alberta,\\
Edmonton, Alberta, Canada T6G 2J1\\
and\\
Departmentof Physics, Technion-Israel Institute of Technology,\\ Haifa
32000, Israel}

\maketitle

\begin{abstract}

Two mode squeezed vacuum states allow  Bell's inequality violation (BIQV)
for all non-vanishing squeezing parameter $(\zeta)$. Maximal violation
occurs at
$\zeta \rightarrow \infty$ when the parity of either component averages to
zero. For a
given entangled {\it two spin}
system BIQV is optimized via orientations of the operators entering the
Bell operator
(cf. S. L. Braunstein, A. Mann and M. Revzen: Phys. Rev. Lett. {\bf68},
3259
(1992)).
We show that for finite $\zeta$
in  continuous variable systems  (and in general whenever the
dimensionality of the subsystems is greater than 2 )
additional parameters are present for optimizing BIQV. Thus
the expectation value of the Bell operator depends, in addition to the
orientation
parameters, on configuration parameters. Optimization of these
configurational parameters leads
to a unique maximal BIQV that depends only on $\zeta.$ The configurational
parameter variation is used to show that BIQV relation to entanglement is,
 even for pure state, not monotonic. 

 \end{abstract}


\newpage

Two mode squeezed vacuum states (TMSV) generated via pulsed non-degenerate 
parametric amplifiers are 
of interest following the pioneering study of Grangier 
et al.~\cite{granier}. Such studies are, 
aside from their intrinsic interest, valuable for clarification of the 
applicability of Bell's 
inequalities for continuous variables in general and the original EPR 
state~\cite{epr} in 
particular. The latter's relevance to the Bell inequality problem was 
stressed originally by 
Bell~\cite{bell}. Such studies pertain to the question whether Quantum 
Mechanics can be 
underpinned with a theory employing local hidden variables. They gained 
particular 
attention with the demonstration, first achieved by Banaszek and 
Wodkiewicz~\cite{wodkiewicz}, 
that negativity  of the Wigner function is not a prerequisite to BIQV. 
In this work we 
demonstrate the presence of configurational parameters  controlling the 
extent of  
BIQV, which are not present in the Bohm-like entangled state (since that
state involves only 
two levels per subsystem, and therefore only orientational parameters are
available for optimization of BIQV.) This is 
particularly relevant for {\it non}-maximally entangled states and should 
be of importance for possible applications of such states. 
  We exhibit two exact solutions for ``local maximal'' 
BIQV for different configurational parameterizations (one was obtained 
earlier in~\cite{Zhang}) 
and give the transformation between them. Our representative of Bells' 
inequalities is the 
CHSH inequality~\cite{Clauser} and the observable is the parity rather 
than the spin. The novel  
configurational dependence introduced herewith should be important when 
such states are employed 
e.g. for delineating a dichotomic variable (even/odd parity in our case) 
as the 
verification involves different observables (exact photon number in one case
 or simply even/odd number of photons in another for the cases 
we consider below). 
BIQV for these states were observed experimentally in several laboratories 
~\cite{mandel,ou} 
within the general study of the problem for continuous variable cases 
\cite{chiao,franson}. 

The state under study, $|\zeta\rangle$ (i.e. $|TMSV\rangle$), involves two 
beams each headed in a 
different direction and delineated by having the operators of the 
first channel 
designated by ${\bf a}=1/\sqrt{2}({\bf q}+i{\bf p})$ and 
${\bf a}^{+}=1/\sqrt{2}({\bf q}-i{\bf p})$ and the second channel by 
${\bf b}=1/\sqrt{2}({\bf q}'+i{\bf p}')$ and 
${\bf b}^{+}=1/\sqrt{2}({\bf q}'-i{\bf p}')$. The TMSV
state, $|\zeta\rangle$, is given by,
\begin{equation}
|\zeta\rangle = S(\zeta) |00\rangle; \;\;\;\;\; S(\zeta)  \equiv\exp[ 
\zeta (a^{+}b^{+} - ab)],
\label{1}
\end{equation}
where $\zeta$ is real.  
The CHSH inequality involves the so-called Bell operator 
${\cal B}$,~\cite{sam}, 
which, for our case, where parity is the dynamical variable, is given via
the following observables:

 (a) The parity operator (the superscript i =1,2
refers to the channel - it is omitted where clarity allows), 
\begin{eqnarray}
{\bf\Pi}_{z} & = & \int_{0}^{\infty}dq \left(|{\cal E}\rangle\langle {\cal 
E}| 
- |{\cal O}\rangle\langle {\cal O}| \right)\; \equiv\; {\bf I}_{E}-{\bf I}_{O},
 \nonumber \\
|{\cal E}\rangle & = & {1 \over \sqrt 2}[ |q\rangle + |-q\rangle ], \quad  
|{\cal O}\rangle = {1 \over \sqrt 2}[ |q\rangle - |-q\rangle ],
\end{eqnarray}
where $|q\rangle$ is the eigenstate of the position operator
and  ${\bf I}_{E}$ and ${\bf I}_{O}$ can be viewed as the identity 
operators in the subspaces of even and odd parity, respectively. We note that
an alternative, equivalent represetation for the latter operators is given 
in the number representation by
\begin{equation}
{\bf I}_{E} =  \sum_{n=0}^{\infty}|2n\rangle\langle 2n|,\;\;\;\;\;\; 
{\bf I}_{O} =  \sum_{n=0}^{\infty} |2n+1\rangle\langle 2n+1|,
\end{equation}
We refer to these two 
representations of ${\bf I}_{E}, {\bf I}_{O}$ given in  Eq.(2) and Eq.(3)
as two configurational representations of the operators. The proof of the 
equivalence of the two configurational representatives for the operator
 ${\bf \Pi}_{z}$ 
 can be easily verified by considering their matrix elements, e.g., with
 respect to the number states.

(b) For the $x$ and $y$ components of $\vec{\bf\Pi}^{(i)}$  we choose
\begin{eqnarray}
{\bf\Pi}_{x}\;\;
& = &\;\; \int_{0}^{\infty}\left(|{\cal E}\rangle\langle 
{\cal O}| + | {\cal O}\rangle\langle{\cal E}|\right)dq\nonumber\\
{\bf\Pi}_{y}\;\;
& = &\;\; i\int_{0}^{\infty}\left(|{\cal O}\rangle\langle 
{\cal E}| - | {\cal E}\rangle\langle {\cal O}|\right)dq.
\label{aa}
\end{eqnarray}
 The components of 
$\vec{\bf\Pi}^{(i)}=({\bf\Pi}^{(i)}_{x},{\bf\Pi}^{(i)}_{y},
{\bf\Pi}^{(i)}_{z})$ ($i=1,2$) satisfy the standard $SU(2)$ algebra, and
the square of each component is equal to the identity operator.
 
With these definitions the Bell operator ${\cal B}$ is given by
\begin{eqnarray}
{\cal B} & = & \vec{n}\cdot\vec{\bf \Pi} 
^{(1)}\otimes\vec{m}\cdot{\bf\vec{\Pi}}^{(2)} 
+\vec{n}'\cdot\vec{\bf \Pi} 
^{(1)}\otimes\vec{m}\cdot{\bf\vec{\Pi}}^{(2)} \nonumber\\
& + & \vec{n}\cdot\vec{\bf \Pi} 
^{(1)}\otimes\vec{m}'\cdot{\bf\vec{\Pi}}^{(2)}
-\vec{n}'\cdot\vec{\bf \Pi} ^{(1)}\otimes\vec{m}'\cdot{\bf\vec{\Pi}}^{(2)},
\label{3}
\end{eqnarray}
and the Bell inequality is, 
\begin{equation}
|\langle {\cal B}\rangle| \leq 2.
\label{2}
\end{equation}
Here $\vec{n},\vec{n}',\vec{m},\vec{m}'$ are unit vectors, specifying 
 the orientational parameters of the first  
and the second channel, respectively. 
In the notation of Eq.(2) (i.e. ${\cal E},{\cal O}$) the state 
$|\zeta\rangle$, Eq.~(\ref{1}), is given by
\begin{equation}
|\zeta\rangle = \int_{0}^{\infty} \int_{0}^{\infty}dqdq' \left[(g_{+} + 
g_{-})|{\cal EE'}\rangle
+(g_{+} - g_{-})|{\cal OO'}\rangle\right],
\end{equation}
with 
$$ 
g_{\pm}(q,q';\zeta) = \langle qq'|S(\pm \zeta)|00\rangle ,
$$ 
giving 
\begin{equation}
g_{\pm} = {1 \over \sqrt\pi} exp\left(-{1 \over 2}[ q^{2} + q'^{2}  
\mp  2qq' \tanh(2\zeta)]\cosh(2\zeta)\right).
\end{equation}
We have directly ${\bf \Pi}_{z}^{(1)}\otimes{\bf \Pi}_{z}^{(2)}|\zeta\rangle = 
|\zeta\rangle$. 


We may now evaluate the Bell operator, Eq.~(\ref{3}),
and choose the angles
(i.e., the unit vectors $\vec{n},\;\vec{n}',\;\vec{m}$ and $\vec{m}'$) 
to {\it maximize} BIQV~\cite{gisin,Zhang},
yielding (we refer to this as the orientational optimization):
\begin{equation}
\langle\zeta|{\cal B}|\zeta\rangle = 2\sqrt {1 + F(\zeta)^{2}},
\label{ggg}
\end{equation}
with $F(\zeta) \equiv 
\langle\zeta|{\bf\Pi}_{x}^{(1)}\otimes{\bf\Pi}^{(2)}_{x}|\zeta\rangle$. 
A straightforward calculation yields
\begin{equation}
F(\zeta) = 2\int_{0}^{\infty}dqdq'\left(g_{+}^{2} - g_{-}^{2}\right) 
= { 2 \over \pi}\arctan\left(\sinh(2\zeta)\right).
\end{equation}
Clearly at $\zeta = 0$ no violation is possible, while for 
$\zeta \rightarrow \infty$ the 
violation attains Cirel'son's~\cite{cirelson} limit. 

To demonstrate the relevance of the configuration  
we now compare two configurational paramerizations: the one studied above with 
the one studied earlier  in~\cite{Zhang} where 
number configurational parametrization were used. We denote by 
$\vec{\bf S}^{(1)}$ and $\vec{\bf S}^{(2)}$
the corresponding operators defined by ~\cite{Zhang},  
they are given in the number state representation by  
${\bf\Pi}_{z}={\bf S}_{z}$ 

\begin{equation} 
{\bf S}_{z} = \sum_{n=0}^{\infty}\left(|2n\rangle\langle 2n| 
- |2n+1\rangle\langle 2n+1|\right) = {\bf I}_{E}-{\bf I}_{O};
\end{equation}
this is the parity operator and it is identical to the one defined in
Eq. (2) (cf. Eq. (3)).The $x$ and $y$ components are given by


\begin{eqnarray}
{\bf S}_{x}\; & = &\;\sum_{n=0}^{\infty}
(|2n\rangle\langle 2n+1|+|2n+1\rangle\langle 2n|)\nonumber\\
{\bf S}_{y}\; & = &\; -i\sum_{n=0}^{\infty}
(|2n\rangle\langle 2n+1|-|2n+1\rangle\langle 2n|).
\label{bb}
\end{eqnarray}

The difference in the value of 
$\langle\zeta|{\bf\Pi}^{(1)}_{x}\otimes{\bf\Pi}^{(2)}_{x}|\zeta\rangle$  
stems from the different choices for ${\bf\Pi}_{x}$ and ${\bf\Pi}_{y}$. 
These are related via a (configurational) unitary transformation 
(commuting with ${\bf\Pi}_{z}$, of course). That is, 
${\bf S}_{\pm}={\bf U}{\bf\Pi}_{\pm}{\bf U}^{\dag}$, where 
${\bf S}_{\pm}=({\bf S}_{x}\pm i{\bf S}_{y})/2$,
${\bf\Pi}_{\pm}=({\bf\Pi}_{x}\pm i{\bf\Pi}_{y})/2$ and
${\bf U}$ is a unitary operator (for example, one can take the 
{\it unitary} operator
${\bf U}={\bf I}_{E}+{\bf S}_{-}{\bf\Pi}_{+}$ or 
${\bf U}={\bf I}_{O}+{\bf S}_{+}{\bf\Pi}_{-}$). 
Thus we may consider these 
transformations as revealing new parameters for optimization of the BIQV 
for the state under study. The configurational parameter dependence implies
(we give an explicit example below) that BIQV in not a monotonic function 
of the entanglement (in our case the latter is parametrized by $\zeta$; 
 $\zeta \rightarrow \infty$ gives maximal entanglement). We first show that 
the choice of  
Chen et al.~\cite{Zhang} leads to maximal violation for all values of 
$\zeta$. 
(Note that $\tanh2\zeta \geq {2 \over 
\pi}\arctan \left(\sinh2\zeta\right)$ 
for all values of $\zeta$,  Fig.1). 
Let us consider ${\bf\Pi}_{x}$ and 
${\bf\Pi}_{y}$ to be arbitrary operators that satisfy the $SU(2)$ 
commutation rules 
with the parity operator ${\bf\Pi}_{z}={\bf I}_{E}-{\bf I}_{O}$ 
as defined above. Thus, the requirement 
${\bf\Pi}_{x}^{2}={\bf\Pi}_{y}^{2}=1$ can be 
written in the form
\begin{equation}
{\bf\Pi}_{+}{\bf\Pi}_{-}={\bf I}_{E}\;\;{\rm 
and}\;\;{\bf\Pi}_{-}{\bf\Pi}_{+}={\bf I}_{O}.
\label{gg}
\end{equation}

Consider now two sets of such operators ${\bf\Pi}_{\pm}^{(i)}$ ($i=1,2$), 
both satisfying
Eq.~(\ref{gg}). Since ${\bf\Pi}_{+}^{(i)}$ transforms even parity functions 
into odd ones, and 
${\bf\Pi}_{-}^{(i)}$ transforms odd functions into even ones, it follows that  
 \begin{equation}
F(\zeta)=\langle\zeta| {\bf\Pi}^{(1)}_{x}\otimes 
{\bf\Pi}^{(2)}_{x}|\zeta\rangle 
= \langle\zeta|{\bf\Pi}^{(1)}_{+}\otimes {\bf\Pi}^{(2)}_{+}|\zeta\rangle 
+\langle\zeta|{\bf\Pi}^{(1)}_{-}\otimes {\bf\Pi}^{(2)}_{-}|\zeta\rangle.
\end{equation}


In the derivation of Eq.~(\ref{ggg}) via orientational parameters optimization
 ~\cite{gisin}, all the azimuthal angles of the 
 vectors 
$\vec{n},\;\vec{n}',\;\vec{m}$ and $\vec{m}'$ in Eq.~(\ref{3}) have been 
set to zero, and we consider the case of
\begin{equation}
\langle\zeta|{\bf\Pi}^{(1)}_{+}\otimes {\bf\Pi}^{(2)}_{+}|\zeta\rangle
=\langle\zeta|{\bf\Pi}^{(1)}_{-}\otimes 
{\bf\Pi}^{(2)}_{-}|\zeta\rangle\;>0.
\label{cond}
\end{equation}  
(Both ${\bf\Pi}_{\pm}$, that follow from Eq.~(\ref{aa}), and ${\bf S}_{\pm}$, 
that follow
from Eq.~(\ref{bb}), satisfy this condition.) In general this requirement, 
Eq. ~(\ref{cond}), is relevant for the orientational parameters optimization
~\cite{gisin} and we assume its validity in our considerations where the
orientational parameters were optimized for fixed configurational ones. 

We now label $\langle 2n|{\bf\Pi}_{+}^{(1)}|2m+1\rangle$ by $U^{1}_{n,m}$ 
and  
$\langle 2n|{\bf\Pi}_{+}^{(2)}|2m+1\rangle$ by $U^{2}_{m,n}$. From 
Eq.~(\ref{gg}) it follows 
that $U^1$ and $U^2$ are unitary matrices. Note also that 
$(U^{1})^{\dag}_{n,m} =\langle 2m|{\bf\Pi}^{(1)}_{-}|2n+1\rangle$, 
and  $(U^{2})^{\dag}_{m,n} =\langle 2m|{\bf\Pi}^{(2)}_{-}|2n+1\rangle$.
In the number representation our state, $|\zeta \rangle$, is given by
\begin{equation}
|\zeta\rangle 
=\frac{1}{\cosh\zeta}\sum_{n=0}^{\infty}(\tanh\zeta)^{n}|nn\rangle.
\end{equation}
We thus further define
the reduced density matrix 
\begin{equation}
\rho_{n,m}\equiv \frac{(\tanh\zeta)^{2n}}{(\cosh\zeta)^{2}}\delta_{n,m}.
\end{equation}
With these definitions we may write 
\begin{eqnarray}
F(\zeta) & = & \frac{1}{2}\sinh(2\zeta)
\left[{\rm Tr}\left(\rho U^{1} \rho U^{2}\right) 
+{\rm Tr}\left(\rho (U^{1})^{\dag} \rho 
(U^{2})^{\dag}\right)\right]\nonumber\\
& = & 
\sinh(2\zeta)
{\rm Tr}\left(\rho U^{1} \rho U^{2}\right), 
\end{eqnarray}
where the last equality follows from the condition~(\ref{cond}).
Hence, we have
\begin{equation}
F(\zeta) \leq \sinh(2\zeta){\rm Tr}\rho ^{2} = \tanh(2\zeta).\;\;\; \rm {QED} 
\end{equation}
 We now give an explicit example of a choice of configurational paraemeters
for which BIQV dependence on $\zeta$ implies BIQV $\rightarrow 0$ with
$\zeta \rightarrow \infty$ while BIQV $\neq 0$ for $\zeta > 0$. i.e.
BIQV is {\it not} a monotonic function of the entanglement even for 
a pure state (note: $\zeta \rightarrow \infty$ implies maximal entanglement):

In this example we take 
${\bf\Pi}_{z}^{(1)}={\bf\Pi}_{z}^{(2)}\equiv {\bf\Pi}_{z}={\bf S}_{z}$ to be 
the parity 
operator as represented in Eq.(2) and ${\bf\Pi}_{x}^{(1)}=
{\bf\Pi}_{x}^{(2)}
\equiv {\bf\Pi}_{x}$ and ${\bf\Pi}_{y}^{(1)}={\bf\Pi}_{y}^{(2)}
\equiv {\bf\Pi}_{y}$ are defined as follows:
\begin{equation}
{\bf\Pi}_{+}^{\dag}={\bf\Pi}_{-}= \frac{{\bf\Pi}_{x}-i{\bf\Pi}_{y}}{2}
=\sum_{n=0}^{\infty}i^{n} |2n+1\rangle\langle 2n|.
\end{equation}
It is simple to check that ${\bf\Pi}_{x}$, ${\bf\Pi}_{y}$ and ${\bf\Pi}_{z}$
satisfy the conventional $SU(2)$ commutation relations. Note that these 
definitions 
corresponds to $U^{1}_{n,m}=U^{2}_{n,m}=(-i)^{n}\delta _{n,m}$. Thus,
\begin{equation}
F(\zeta)=\sinh(2\zeta)
{\rm Tr}\left(\rho U^{1} \rho U^{2}\right)=\frac{\sinh(2\zeta)}{\cosh^{4}\zeta}
\sum_{n=0}^{\infty}(-1)^{n}(\tanh\zeta)^{4n}=\sinh(2\zeta)
\frac{(1-\tanh^{2}\zeta)^{2}}
{1+\tanh^{4}\zeta}.
\end{equation}
Note that $F(\zeta =0)=F(\zeta\rightarrow\infty)=0$ while $F(\zeta) > 0$ 
for finite (non-zero) values of $\zeta$. 



The dichotomic ($\pm 1$) parity operator is studied for the continuous 
variable two 
channel system of two mode squeezed vacuum state. The maximal BIQV for the 
Bell operator is 
demonstrated explicitly for two cases. 
The violation of the inequality is shown to depend on the choice of these
components for non-maximal entanglement thus revealing configurational 
parameters whose 
choice allows optimization of the violation. We show that the particular 
choice, 
selected in~\cite{Zhang} attains this optimization and that the dependence 
on these new (configurational) parameters implies that Bell's inequality   
violation is {\it not} a monotonic function of the entanglement. 
\section*{Acknowledgment}
We thank R.~Teshima for his help. 
G.G. research is supported by the Killam Trust;       
F.K. research is supported by NSERC.

\begin{figure}[htb]      
\caption{Variation of $F(\zeta)$ (see Eq.~(\ref{ggg})) with $\zeta$ for two
configurational schemes.}
\end{figure}

\end{document}